\documentclass[useAMS, usenatbib, a4paper]{mn2e}
\usepackage{aas_macro, amsmath, amssymb, color, graphicx, multirow, setspace}
\begin{document}

\title[TriAnd and its Siblings]{TriAnd and its Siblings: Satellites of Satellites in the Milky Way Halo}

\author[A. J. Deason et al.]{A. J. Deason$^{1}$\thanks{E-mail: alis@ucolick.org}\thanks{Hubble Fellow}, V. Belokurov$^{2}$, K. M. Hamren$^{1}$, S. E. Koposov$^{2,3}$, K. M. Gilbert$^{4}$, \newauthor R. L. Beaton$^{5}$, C. E. Dorman$^{1}$, P. Guhathakurta$^{1}$, S. R. Majewski$^{5}$, E. C. Cunningham$^{1}$\\
$^{1}${Department of Astronomy and Astrophysics, University of California Santa Cruz, Santa Cruz, CA 95064, USA}\\
$^{2}${Institute of Astronomy, University of Cambridge, Madingley Road, Cambridge, CB3 0HA, UK}\\
$^{3}${Moscow MV Lomonosov State University, Sternberg Astronomical Institute, Moscow 119992, Russia}\\
$^{4}${Space Telescope Science Institute, 3700 San Martin Drive, Baltimore, MD 21218, USA}\\
$^{5}${Department of Astronomy, University of Virginia, Charlottesville, VA, USA}}

\date{\today}

\pagerange{\pageref{firstpage}--\pageref{lastpage}} \pubyear{2014}

\maketitle

\label{firstpage}

\begin{abstract}
We explore the Triangulum-Andromeda (TriAnd) overdensity in the SPLASH (Spectroscopic and Photometric Landscape of Andromeda's Stellar Halo) and SEGUE (the Sloan Extension for Galactic Understanding and Exploration) spectroscopic surveys. Milky Way main sequence turn-off stars in the SPLASH survey reveal that the TriAnd overdensity and the recently discovered PAndAS stream (\citealt{martin14}) share a common heliocentric distance ($D \sim 20$ kpc), position on the sky, and line-of-sight velocity ($V_{\rm GSR} \sim 50$ km s$^{-1}$). Similarly, A-type, giant, and main sequence turn-off stars selected from the SEGUE survey in the vicinity of the Segue 2 satellite show that TriAnd is prevalent in these fields, with a velocity and distance similar to Segue 2. The coincidence of the PAndAS stream and Segue 2 satellite in positional and velocity space to TriAnd suggests that these substructures are all associated, and may be a fossil record of group-infall onto the Milky Way halo. In this scenario, the Segue 2 satellite and PAndAS stream are ``satellites of satellites'', and the large, metal-rich TriAnd overdensity is the remains of the group central. 
\end{abstract}


\section{Introduction}
In the modern cold dark matter cosmology ($\Lambda$CDM), dark matter halos form hierarchically, and the presence of ``subhalos'' or ``sub-subhalos'' is naturally predicted. Dark matter subhalos are likened to the satellite galaxies that orbit the Milky Way (MW) galaxy, and the agreement/disagreement between the observed satellite population and the predictions from simulations are well documented (e.g, \citealt{flores94}; \citealt{moore94}; \citealt{klypin99}; \citealt{moore99}). However, the population of sub-subhalos (or satellites of satellites) is less well studied. In essence, smaller constituents should be associated with the satellite population today and/or the destroyed remnants of accreted satellites in the stellar halo. It remains to be seen what fraction of these sub-subhalos have survived to the present day, or even if we are able to observe them (i.e. if they are too faint, or completely ``dark'').

Observational evidence for associations between satellites and/or streams increased dramatically following the discovery of very low luminosity galaxies ($L \lesssim 10^5 L_\odot$) in the MW (e.g, \citealt{willman05}; \citealt{belokurov06b, belokurov07a}). Amongst the faintest of these, curious associations in both position and velocity space with other satellites or streams have been found. For example, \cite{newberg10} showed that the Orphan stellar stream (\citealt{belokurov06a}; \citealt{grillmair06}; \citealt{belokurov07b}) has a similar distance and velocity as the Segue 1 satellite, and is only $\sim 2$ deg away on the sky. The narrow width of the Orphan stream ($\sim 1$ deg) indicates that Segue 1 is probably not the progenitor of the stream, but it is likely that the two are dynamically associated. The proximity of the Bo{\"o}tes II satellite, in both position and velocity space, with the Sagittarius stream led \cite{koch09} to suggest that this ultra-faint may have been stripped from the more massive Sgr dwarf. Similarly, \cite{belokurov09} noted the nearness of the Segue 2 satellite (perhaps the ``least massive galaxy'' known; \citealt{kirby13a}) to the elusive TriAnd overdensity (see below), which again suggests that a very faint satellite may be associated with a more massive structure. Finally, the two ultra-faints,  Leo IV and Leo V, discovered in quick succession by \citet{belokurov07a,belokurov08} are separated by $\sim 3$ deg on the sky, with small offsets in both distance and velocity ($\sim 20$ kpc, $\sim 40$ km s$^{-1}$); \cite{belokurov08} dub Leo V as ``a companion of a companion of the Milky Way galaxy''.

The enhanced evidence for satellite-satellite associations from the lowest mass galaxies is perhaps unsurprising. Wetzel, Deason \& Garrison-Kimmel (2014, in prep.) recently showed that a significant fraction ($\sim 30$ percent) of low mass subhalos ($M_{\rm star} \lesssim 10^5M_\odot$) likely fell into a MW-type host as a satellite of a more massive subhalo. These results suggest that the most likely place to find associations between satellites and streams today is in the vicinity of the most massive substructures in the Galaxy, which may have hosted less massive sub-subhalos at the time of infall.

Among the various streams/overdensities identified in the MW halo, there are five ``big ones'' that stand out the most: the Sagittarius stream (Sgr, \citealt{ibata94}; \citealt{yanny00}; \citealt{majewski03}), the Monoceros Ring\footnote{Note that it has also been claimed that the Monoceros Ring is the signature of disk flaring rather than the signature of a disrupted dwarf galaxy (e.g, \citealt{ibata03}; \citealt{momany06}).} (\citealt{newberg02}; \citealt{ibata03}), the Hercules Aquila cloud (\citealt{belokurov07c}; \citealt{simion14}), the Virgo overdensity (\citealt{vivas02}; \citealt{juric08}; \citealt{bonaca12a}), and the Triangulum-Andromeda stream (TriAnd, \citealt{majewski04}; \citealt{rocha04}; \citealt{sheffield14}). It is likely that these are the remnants of fairly massive accretion events, and they each span several hundreds of degrees on the sky. Most of these structures have been studied extensively, largely thanks to the wide sky coverage of the Sloan Digital Sky Survey (SDSS). However, TriAnd, which is the least well-known, and perhaps the smallest of the five, has evaded significant scrutiny in the literature.

The TriAnd overdensity, so-called from its discovery in the constellations of Triangulum and Andromeda (\citealt{majewski04}; \citealt{rocha04}), is a large, diffuse structure that lies at a heliocentric distance of $10 \lesssim D/\mathrm{kpc} \lesssim 30$. The full extent of this overdensity is poorly known, it subtends at least $50^\circ \times 60^\circ$ on the sky, but is located in a region of the sky that has not been completely covered by large surveys such as SDSS. As a result, the TriAnd overdensity has received relatively little attention. In fact, most of the work on TriAnd has resulted from serendipitous observations focused on the M31 galaxy (\citealt{majewski04}; \citealt{martin07}; \citealt{martin14}). The nature of TriAnd has been debated in the literature; \cite{rocha04} propose that is is either a bound core of a very dark-matter dominated dwarf or a portion of a tidal stream, while \cite{johnston12} suggest that TriAnd is similar to ``cloud-like'' debris from a disrupted dwarf galaxy (see also \citealt{deason13}; \citealt{sheffield14}).

TriAnd is not the only MW substructure discovered from M31 surveys;  \cite{martin14} recently discovered a new faint stellar stream located at $D\sim20$ kpc in their photometric survey of M31. Most surveys of the MW halo tend to avoid regions close to the Galactic disk, so this detection suggests there may be several low latitude streams awaiting discovery. The stream is dynamically cold ($\sigma < 7.1$ km s$^{-1}$) and has a width of 300-650 pc, which lead \cite{martin14} to argue for a dwarf-galaxy-accretion origin. 

In this contribution, we study the TriAnd overdensity in two large spectroscopic surveys, SPLASH and SEGUE, and argue for the association of a stream (the newly discovered PAndAS stream) and a satellite (Segue 2) with this more massive structure. The paper is arranged as follows. Section 2 briefly describes the spectroscopic surveys used in this work. In Sections 3 and 4 we use SPLASH and SEGUE to identify stars likely belonging to the TriAnd overdensity along the line-of-sight towards M31, and in the vicinity of the Segue 2 satellite. We discuss the implications of our results in Section 5, and we summarize our main conclusions in Section 6.

\section{Spectroscopic Data}
\subsection{Spectroscopic and Photometric Landscape of Andromeda's Stellar Halo (SPLASH)}
The SPLASH data that we use in this work is identical to the data used by \cite{gilbert12} to characterize the M31 stellar halo. Here, we provide a brief summary of the observations but refer the interested reader to earlier SPLASH papers for more details (\citealt{guhathakurta06}; \citealt{gilbert12}).

The photometric catalogs used to target M31 red giant branch (RGB) stars for spectroscopic follow-up primarily came from imaging observations obtained with the Mosaic Camera on the Kitt Peak National Observatory (KPNO) 4-m Mayall telescope. Observations were made in the Washington system ($M$, $T_2$, $DD051$), which can be used to efficiently separate giant stars in M31 from foreground MW stars (\citealt{majewski00}). Additional photometry was compiled using $V$ and $I$ (or $g'$ and $i'$) filters obtained from observations with the  William Herschel Telescope, the Suprime-Cam instrument on the Subaru Telescope, and the MegaCam instrument on the 3.6-m CFHT.

\begin{figure}
  \centering
  \includegraphics[width=8.8cm, height=6.6cm]{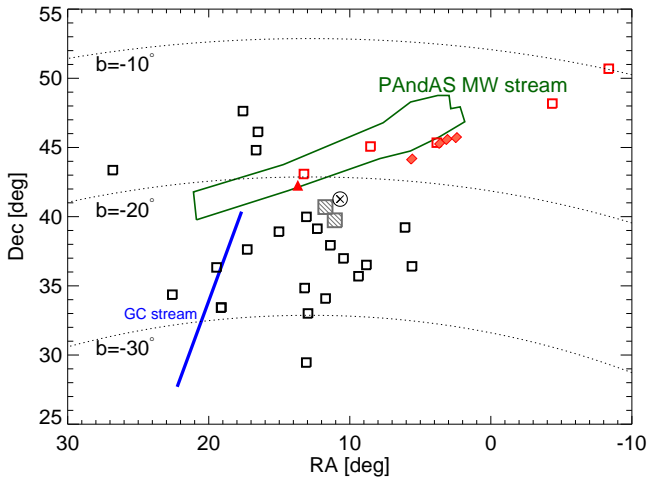}
  \caption[]{Locations of SPLASH fields with spectroscopic measurements included in \cite{gilbert12}. We show the footprints of the KPNO/Mosaic (open squares), CFHT/MegaCam (line-filled grey squares), and Subaru/Suprime-Cam (tilted filled rectangles) images used to target stars for spectroscopic follow-up. The location of the PAndAS stream is indicated with the green region, and fields coincident with the PAndAS stream are highlighted in red. The red triangle indicates the field where \cite{martin14} measured radial velocities for the PAndAS stream stars. The center of M31 is indicated with a cross symbol, and the dotted lines show lines of constant Galactic latitude. The location of the recently discovered globular cluster stream by \cite{bonaca12b} is shown with the blue line. The TriAnd overdensity likely fills most of the region shown(cf. Figure 2 in \citealt{martin14}).}
\label{fig:splash_fields}
\end{figure}

The M31 fields span a large range in position angle and projected distance from the center of M31 (see Fig. \ref{fig:splash_fields}). The above photometric catalogs were used to obtain spectroscopic observations of M31 stars using the DEIMOS spectrograph on the Keck~II telescope. M31 stars were given higher priority for inclusion on the slit mask, but lower priority filler targets, such as MW foreground stars, were often included on the masks due to the paucity of M31 RGB stars (especially in the low density, outer fields of M31). Spectra were obtained over nine observing seasons (Fall 2002\,--\,2010), with typical integration times of 1 hr per mask. The 1200~line~mm$^{-1}$ grating was used for all observations, and the typical wavelength range of the spectra is 6450\,--\,9150\AA.

The reduction of the spectra using the {\tt spec2d} and {\tt spec1d} software \citep{cooper12,newman13} is described in more detail in \cite{gilbert12}. The resulting reduced spectra from 108 masks yielded successful velocity measurements for over 5800 stellar spectra. In this work we are only interested in the MW foreground spectra, specifically the turn-off stars in the MW halo (see \S\ref{sec:splash_trash}).

\subsection{Sloan Extension for Galactic Understanding and Exploration (SEGUE)}

The SDSS (\citealt{york00}) is an imaging and spectroscopic survey covering over one quarter of the sky. A dedicated 2.5m telescope at Apache Point Observatory, New Mexico is used to obtain $ugriz$ imaging (\citealt{fukugita96}) and moderate-resolution (R $\sim 2000$) spectra. SEGUE (including both SEGUE-1 and -2), a key project executed during SDSS-II and SDSS-III, has obtained spectra for $\sim 360, 000$ stars in the MW, and was designed to explore the kinematics and populations of our Galaxy and its halo (\citealt{yanny09}). 

The wavelength coverage of the SEGUE survey spans 3900\AA\ to 9000 \AA\ and targets fainter MW stars ($14.0 < g < 20.4$) of a wide variety of spectral types. The stellar populations include main-sequence stars, hot A-type stars, and evolved giant stars. The SEGUE spectra are clustered in ``pencil-beam'' regions spaced over a significant portion of the sky ($\sim 2800$ deg$^{2}$).

The SDSS Data Release 9 provides estimates of T$\rm _{\rm eff}$, log g, [Fe/H] and [$\alpha$/Fe] from an updated and improved version of the SEGUE Stellar Parameter Pipeline (SSPP, e.g. \citealt{lee08}). Typical uncertainties in the derived parameters are $\sim 4$ km s$^{-1}$ for line-of-sight velocities, and $\sim 0.2$ dex in [Fe/H] metallicity.

\section{TriAnd and a Skinny Friend in the SPLASH ``Trash''}
\label{sec:splash_trash}
In this section, we analyze the TriAnd overdensity using the MW foreground contamination from the SPLASH survey.

\begin{figure}
    \centering
    \includegraphics[width=8.8cm, height=4.4cm]{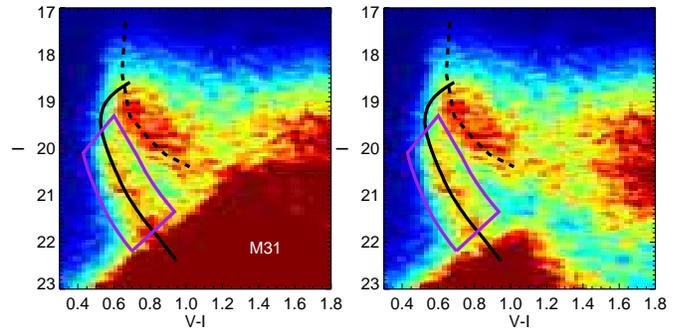}
    \caption[]{Colour-magnitude diagram for stars in the M31 fields. All stars are shown in the left-hand panel. Stars with a relatively low probability of being a giant ($g_{\rm prob} < 0.5$; see \citealt{ostheimer03}) are shown in the right-hand panel. The dashed line indicates the apparent Monoceros Ring ridgeline given by \cite{ibata03}. We note, however, that this structure could also belong to the (thick) disk population. The solid line shows a Padova isochrone with age $T=10$ Gyr and metallicity [Fe/H]=$-1.5$ shifted to $D=17$ kpc. In the following analysis of the SPLASH data, we select stars within the purple selection box as our foreground MW/TriAnd sample.}
\label{fig:washington_cmd}
\end{figure}

In Fig. \ref{fig:splash_fields} we show the spectroscopic footprint of the SPLASH survey fields used in this work. The open squares highlight fields that targeted M31 stars using Washington photometry filters $M, T_2$, and $DD051$ to distinguish between dwarfs and giants (see \citealt{majewski00}). Additional fields that used CFHT/MegaCam or Subaru/Suprime-Cam images to select targets are shown with the line-filled squares and tilted filled rectangles, respectively. The location of the recently discovered PAndAS stream is shown in green (\citealt{martin14}). Several of the SPLASH fields coincide with this structure (highlighted in red), and we explore these fields in more detail below.

\begin{figure*}
  \centering
  \begin{minipage}{0.6\linewidth}
    \centering
    \includegraphics[width=11cm, height=5.5cm]{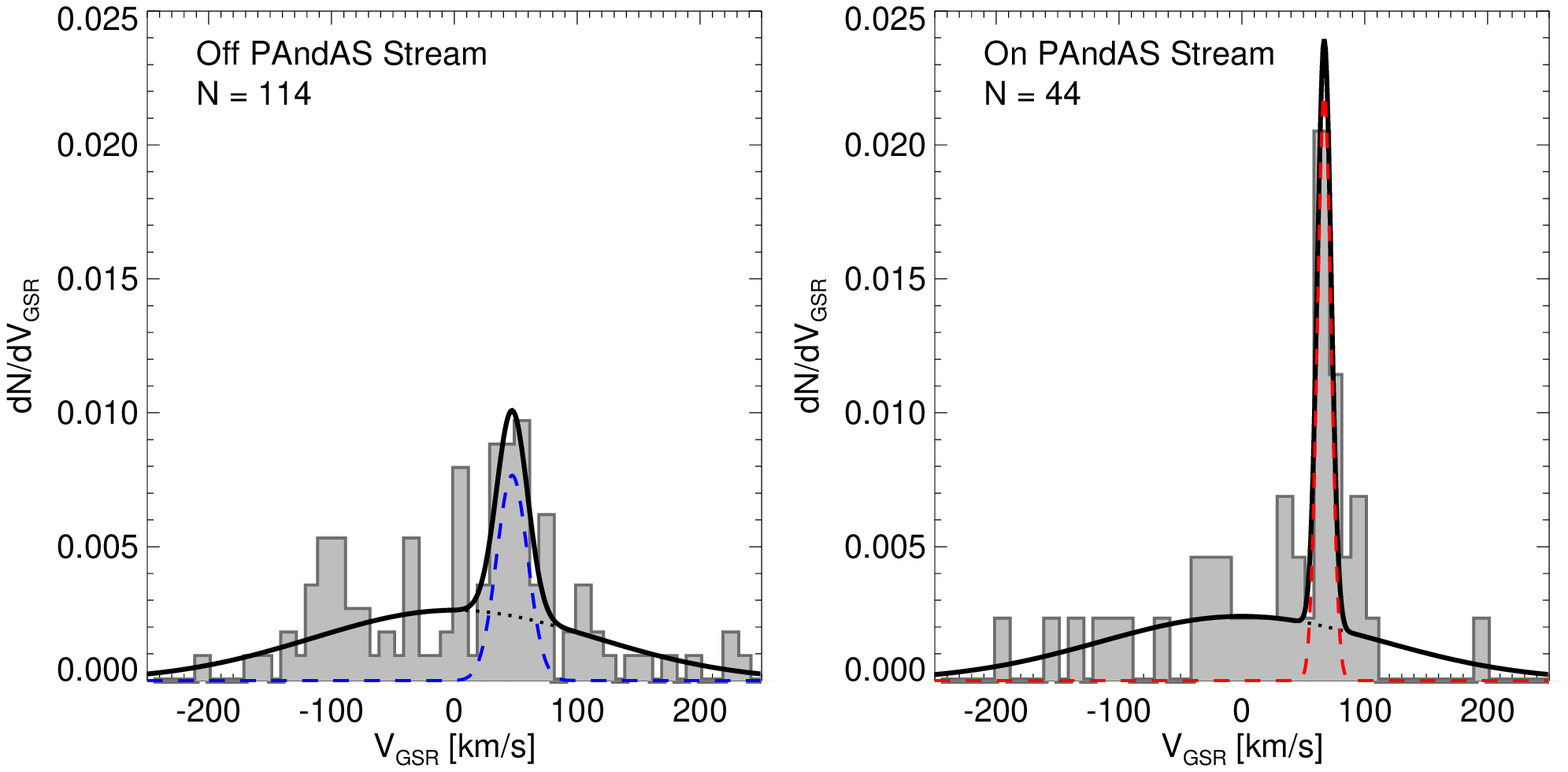}
  \end{minipage}\hfill
  \begin{minipage}{0.37\linewidth}
    \centering
    \includegraphics[width=6.875cm, height=5.5cm]{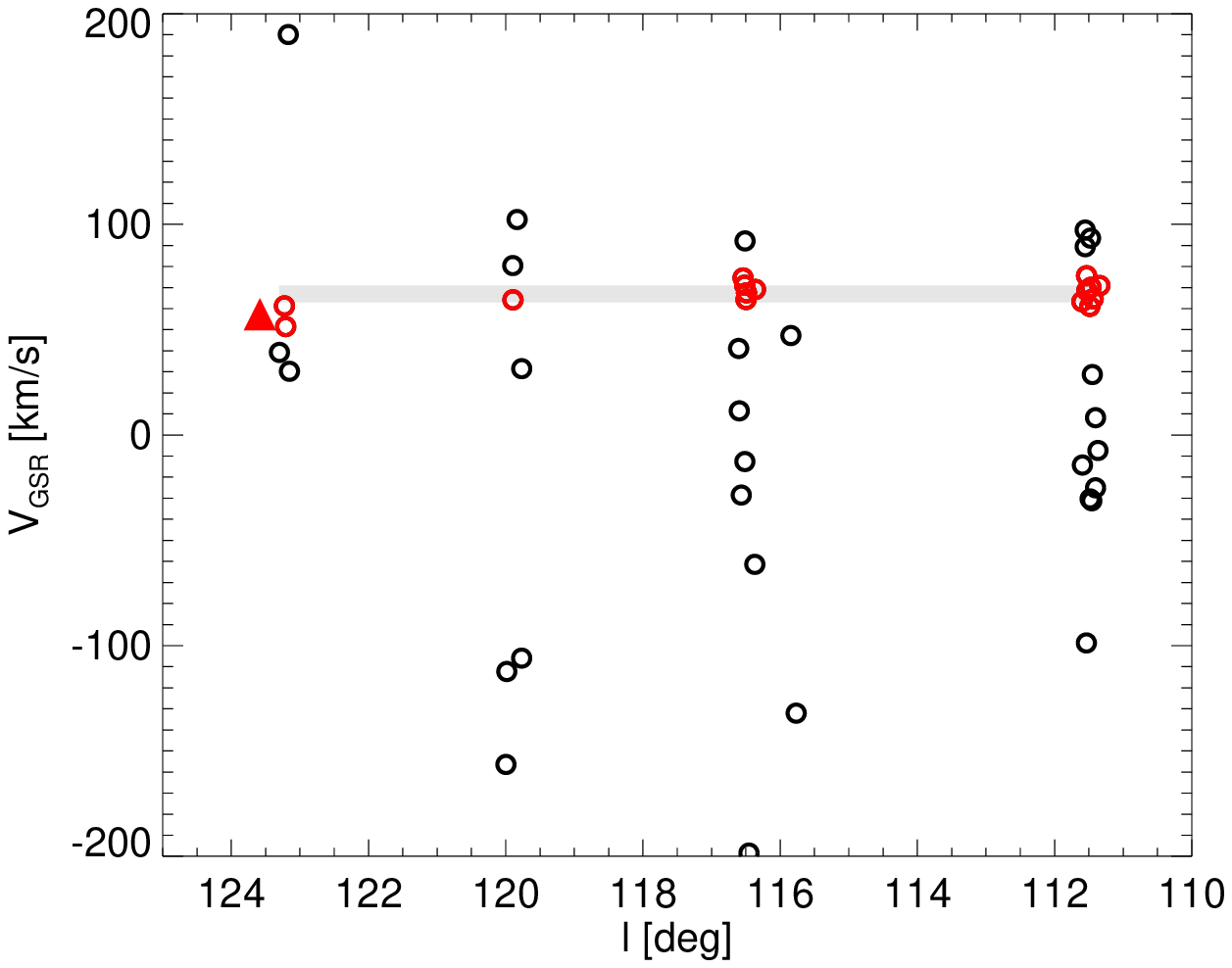}
  \end{minipage}
\caption[]{\textit{Left and middle panels:} Velocity distributions of stars in the TriAnd CMD selection box. Stars in fields ``off'' and ``on'' the PAndAS stream are shown in the left and middle panels, respectively. The maximum likelihood velocity distributions for MW halo and substructure components are shown with the dotted black and dashed (blue/red) lines. The cold component in the fields not coincident with the PAndAS stream (``off'' fields) is TriAnd. \textit{Right panel:} The line-of-sight Galactocentric velocities of stars in the PAndAS stream fields as a function of Galactic longitude. Red circle symbols indicate stars that likely belong to the PAndAS stream ($p_{\rm stream} > 0.5$). Note that in computing $p_{\rm stream}$ we assume a constant density of stars with Galactic longitude. The complicated selection function used to target these stars means that it is difficult to draw conclusions from the varying number of stream stars from field-to-field. The red triangle shows the average velocity for the PAndAS stream found by \cite{martin14}. There is tentative evidence for a velocity gradient along the PAndAS stream with decreasing $l$, but the number of stars at high Galactic longitude is very low.}
\label{fig:splash_vel}
\end{figure*}
Fig. \ref{fig:washington_cmd} shows the $V, I$ colour-magnitude diagram (CMD) for stars in the M31 fields with available Washington photometry. The Washington filters have been converted to Johnson-Cousins bandpasses using the relations given in \cite{majewski00}. For the SPLASH survey, M31 giant stars were given highest priority on the DEIMOS masks, but MW foreground targets were also included to fill the mask space. In fact, for the sparsely populated halo fields targeted by \cite{gilbert12}, the MW foreground stars often outnumber the M31 giants. The right-hand panel excludes high probability giants from the CMD; bright ($I\sim20$) MW main sequence turn-off (MSTO) stars are generally well separated from the M31 giants in CMD space. Two prominent foreground features are visible in the CMD. The feature visible at brighter magnitudes  ($I \sim 19$) could be the Monoceros Ring (\citealt{newberg02}; \citealt{ibata03}), or an extension of the thick disk population. For illustration, we indicate the ridgeline of this feature given by \cite{ibata03}. At fainter magnitudes a second prominent feature is apparent. This is the TriAnd overdensity, which was first noticed in photometric surveys towards M31 by \cite{majewski04}. The solid line shows a Padova isochrone (\citealt{bressan12}) with age $T=10$ Gyr and metallicity [Fe/H]=$-1.5$ shifted to $D=17$ kpc. This isochrone matches the observed distribution in CMD space very well. In the following analysis of the SPLASH data, we select ($N=158$) stars within the purple selection box as our foreground MW/TriAnd sample.  

\subsection{Velocity Structure of Milky Way Populations}
\label{sec:splash_vel}

The Galactocentric velocity distributions of stars in the MW MSTO star selection box are shown in the left-hand panels of Fig. \ref{fig:splash_vel}. Observed heliocentric velocities are converted to Galactocentric ones ($V_{\rm GSR}$) by assuming a circular speed of 235 km s$^{-1}$ at the position of the Sun ($R_0 = 8.5$ kpc) with a solar peculiar motion ($U, V, W$)=(11.1, 12.24, 7.25) km s$^{-1}$ (\citealt{schonrich10}). Here, $U$ is directed toward the Galactic center, $V$ is positive in the direction of Galactic rotation and $W$ is positive towards the North Galactic Pole. The two panels show the velocity distributions ``on'' and ``off'' the PAndAS stream. In both panels there is an excess of stars with velocities at $V_{\rm GSR} \sim 50$ km s$^{-1}$. We fit a two-component Gaussian profile to the velocity distribution. This includes a fixed MW halo population with $V_{\rm halo}=0$ km s$^{-1}$ and $\sigma_{\rm halo}=115$ km s$^{-1}$ (e.g, \citealt{xue08}; \citealt{brown10}), and a cold component with mean velocity and dispersion as free parameters. We use a maximum likelihood algorithm to estimate the best fit Gaussian parameters for the cold component, and the normalizations for both components are computed iteratively for each set of parameters ensuring that the total number of objects is correct. The maximum likelihood parameters are listed in Table \ref{tab:like}.

The Galactocentric velocity of the structures both on and off the PAndAS stream fields are very similar to the velocity of TriAnd found by \cite{rocha04} using M-giants (see Fig. \ref{fig:master}). The velocity dispersion of the TriAnd feature is hotter than the PAndAS stream. The PAndAS stream has a very narrow velocity dispersion ($\sigma \sim 4 $ km s$^{-1}$), in good agreement with the findings of \cite{martin14}. The proximity of this ``skinny'' structure to the TriAnd overdensity, in both positional space (i.e. distance and position on the sky) and line-of-sight velocity,  strongly suggests that they are associated. Note that a possible association between TriAnd and the PAndAS stream is also mentioned by \cite{sheffield14}. We discuss the relation between these stellar structures further in \S \ref{sec:diss}.

The velocities of the PAndAS stream stars as a function of Galactic longitude are shown in the right-hand panel of Fig. \ref{fig:splash_vel}. Based on the Gaussian velocity distribution for the cold component, we can estimate the probability of stream membership, i.e.:
\begin{eqnarray}
\label{eq:prob}
p_{\rm sub}=\frac{f_{\rm sub} G_{\rm sub}}{f_{\rm sub} G_{\rm sub}+f_{\rm halo} G_{\rm halo}} \nonumber\\
G=\frac{1}{\sqrt{2 \pi \sigma}}\mathrm{exp}\left(-(v-\langle v\rangle)^2/2 \sigma^2\right)
\end{eqnarray}
where, $p_{\rm sub}$ is the probability for the relevant substructure (in this case the stream), $G$ is a Gaussian velocity distribution, and $f$ is the fractional contribution of the substructure/halo components.
The circles highlighted in red show those stars with $p_{\rm stream} > 0.5$. There is tentative evidence for a velocity gradient along the PAndAS stream with decreasing $\ell$, but the number of stars at high Galactic longitude is very low.

\subsection{Relative Metallicity of Milky Way Populations}

\begin{figure}
\begin{center}
 \includegraphics[width=8cm, height=6.4cm]{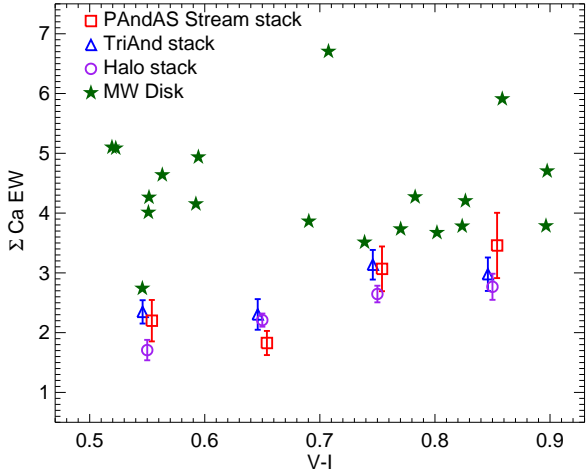}
\caption[]{Calcium triplet equivalent widths as a function of V-I colour. The green stars are high signal-to-noise MW disk stars. The stacked values for the PAndAS stream, TriAnd and field MW halo stars are shown with the red squares, blue triangles and purple circles, respectively. The bins in V-I colour are 0.1 dex wide, and the central values for the of the bin are shown for the x-axis values. The symbols are shown slightly offset on the x-axis for clarity. The average equivalent widths for the stack, and associated error in the mean, are shown with the y-axis values. We find no significant difference in metallicity between the PAndAS stream, TriAnd and field halo stars based on the weak Ca \textsc{ii} lines.}
\label{fig:caew}
\end{center}
\end{figure}

We now consider if there is any significant metallicity difference between the TriAnd feature and the PAndAS stream. The SPLASH spectra cover the wavelength range of the Ca \textsc{ii} triplet ($\lambda 8500$ \AA), which was used in previous work to estimate the metallicity of M31 red giant branch stars (e.g. \citealt{gilbert06}; \citealt{guhathakurta06}; \citealt{ho14}). For the bluer, main-sequence stars considered here, the Ca \textsc{ii} lines are much weaker than for redder stars and the signal-to-noise of the selected spectra are relatively low (S/N $\sim 5$ per \AA\ ). Thus, in order to maximize any apparent metallicity signal in the data, we stack together spectra in $V-I$ colour-bins for each of the different MW substructures. We consider the high probability members of the TriAnd overdensity (with $P_{\rm triand} > 0.5$), PAndAS stream ($P_{\rm stream} > 0.5$) and field halo ($P_{\rm halo} > 0.5$). This corresponds to $N=36 ,15$ and $107$ objects for each of these populations, respectively. 

Fig. \ref{fig:caew} shows the effective Ca \textsc{ii} equivalent width (EW) for stacked SPLASH spectra in the field halo, the PAndAS stream and TriAnd. Only two of the Ca \textsc{ii} triplet lines are used as the line at $\lambda$8498 \AA\ is very weak in the low S/N, hot spectra of the turn-off stars. In addition, it occasionally landed in the inter-CCD gap of the DEIMOS detector for some of the observing runs (\citealt{gilbert06}).  The EW of the Ca \textsc{ii} ($\Sigma \mathrm{Ca}$) triplet was calculated using a linear combination of the EWs of the remaining two lines that has been shown to maximize the S/N of the feature (\citealt{rutledge97}): $\Sigma \mathrm{Ca}=1.0 \mathrm{EW} (\lambda8542$ \AA ) $+ 0.6 \mathrm{EW}(\lambda8662$ \AA ). We note that we use the Ca EW as a proxy for metallicity, but do not enforce a calibration between Ca EW and [Fe/H] as this relationship has (to our knowledge) only been optimized for giant stars.

We find no significant difference between the three populations. For comparison, high S/N spectra for (likely) bright MW disk stars (selected from the same SPLASH fields with $I < 19.5$, $V_{\rm GSR} > 100$ km s$^{-1}$) are shown with the green star symbols. These stars, which are presumably more metal-rich,  have systematically higher Ca \textsc{ii} EWs than the halo populations. This suggests that the weak Ca \textsc{ii} lines are able to differentiate between significant metallicity differences for blue ($V-I < 1.0$) stars. However, it is unclear whether these weak Ca \textsc{ii} lines are able to distinguish between modest metallicity differences for the (metal-poorer) halo populations. We defer a more detailed study of the metallicities of the TriAnd and PAndAS stream structures in the SPLASH survey to future work. 

\section{TriAnd in SDSS/SEGUE: Proximity to the Segue 2 Satellite} 

\begin{figure*}
  \centering
  \begin{minipage}{0.49\linewidth}
    \centering
    \includegraphics[width=8.8cm, height=7.04cm]{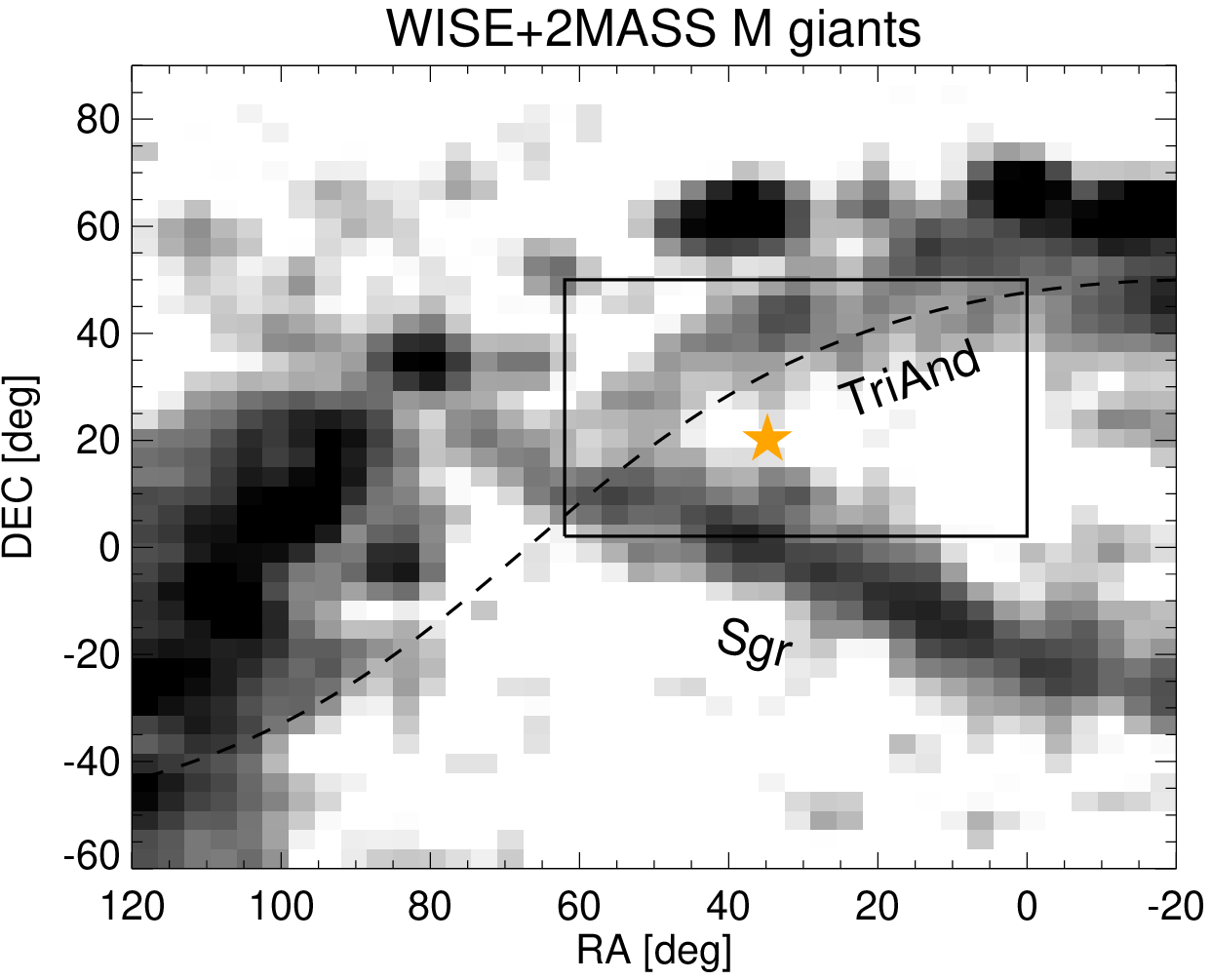}
  \end{minipage}\hfill
  \begin{minipage}{0.49\linewidth}
    \centering
    \includegraphics[width=8.8cm, height=7.04cm]{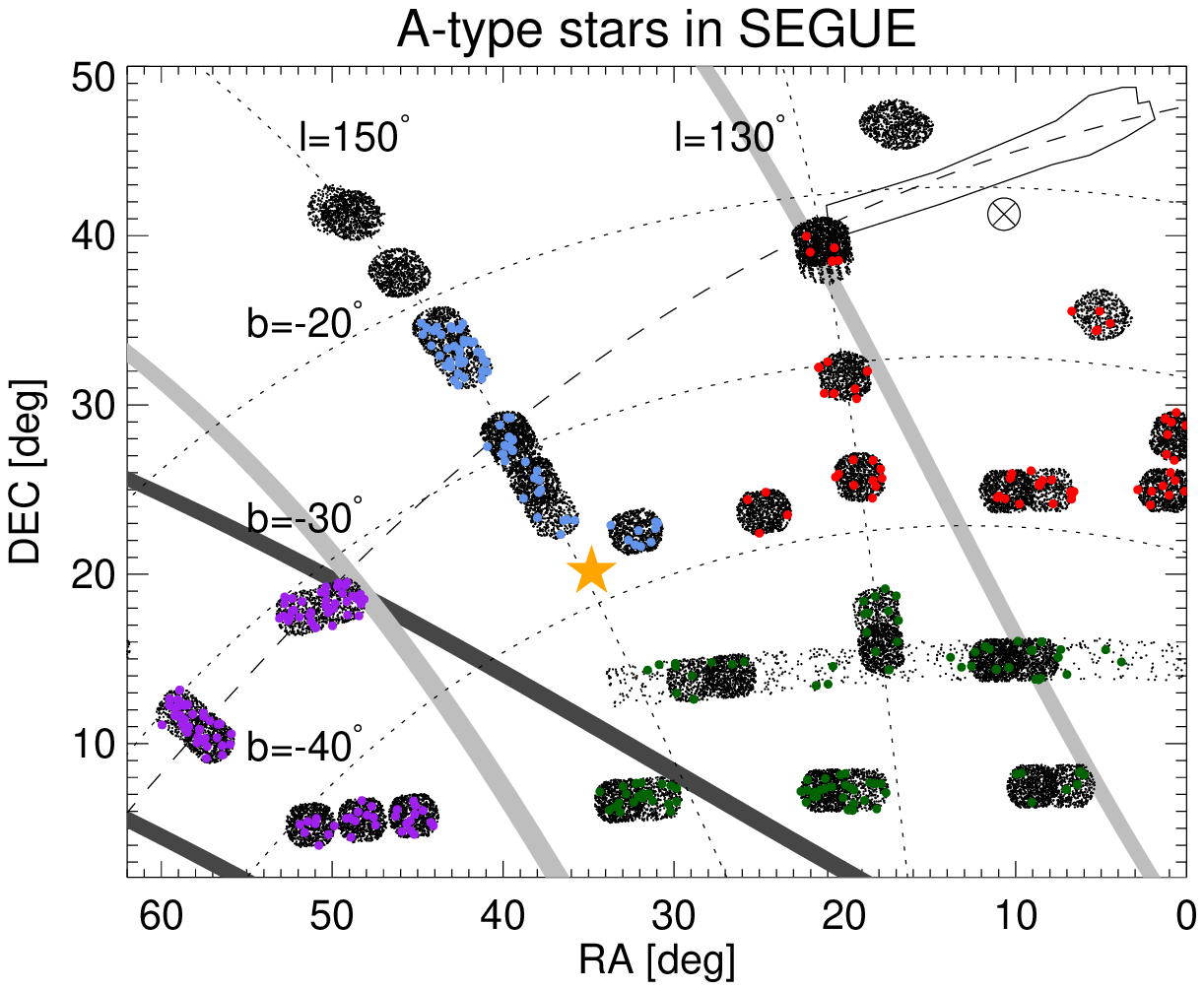}
  \end{minipage}
\caption[]{\textit{Left panel:} Density of M giants selected from WISE and 2MASS photometry (Koposov et al. 2014, in prep.) in equatorial coordinates. The black box indicates the region where we analyze TriAnd in the SDSS/SEGUE spectroscopic fields. The dashed line illustrates a great circle through the PAndAS stream. \textit{Right panel:} The SEGUE spectroscopic footprint in equatorial coordinates.  Fields close to the Segue 2 satellite are shown, and the location of the satellite is indicated by the orange star symbol. A-type stars (mainly blue stragglers) selected in these fields are highlighted in blue, red, purple and green according to their equatorial position on the sky. The location of the PAndAS stream and the center of M31 are highlighted in the top right-hand corner. Lines of constant Galactic latitude/longitude are shown with dotted lines. The approximate tracks of Sgr and Cetus are shown with the shaded dark and light grey shaded regions, respectively (\citealt{belokurov14}).}
\label{fig:radec_segue}
\end{figure*}

\begin{figure*}
    \centering
    \includegraphics[width=14cm, height=10.5cm]{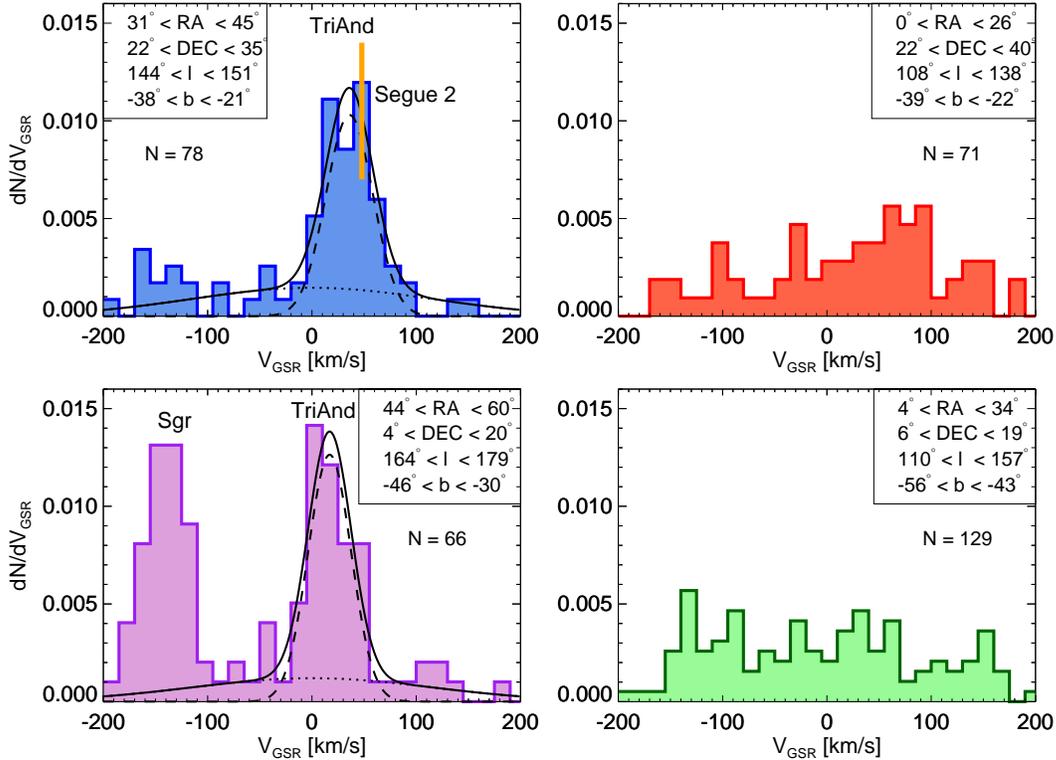}
\caption[]{The Galactocentric velocity distributions of A-type stars located in the selected SEGUE fields. The fields coincident with a great circle through the PAndAS stream, show a significant excess at $V_{\rm GSR} \sim 30$ km s$^{-1}$, which is coincident with the velocity of the TriAnd overdensity and the Segue 2 satellite. The solid line shows a two-component Gaussian fit to the velocity distribution, where the field halo velocity distribution (dotted line) is fixed with $V_{\rm halo} =0$ km s$^{-1}$ and $\sigma_{\rm halo}=115$ km s$^{-1}$. The TriAnd overdensity is less prominent in the fields at lower RA. Velocity signatures associated with Sgr are also highlighted. The velocity distributions of TriAnd and Sgr in the bottom left panel have similar dispersions ($\sim 20$ km s$^{-1}$), suggesting that these may be similar mass structures.}
\label{fig:vlos_segue}
\end{figure*}

In the Segue 2 discovery paper, \cite{belokurov09} noted the proximity of this satellite to TriAnd. Inspired by this possible connection, we now investigate the TriAnd overdensity using A-type, giant and MSTO stars in the vicinity of the Segue 2 satellite.

In the left panel of Fig. \ref{fig:radec_segue} we show the (large-scale) distribution of M giants selected from WISE and 2MASS photometry (Koposov et al. 2014, in prep.). M giants in the approximate distance range $15 \lesssim D/\mathrm{kpc} \lesssim 30$, are selected with the following colour and magnitude cuts:

\begin{eqnarray}
-0.2 < W_1-W_2 < -0.07 \\ \nonumber
0.9 < J-K_s < 1.3 \\ \nonumber
11 < K_s < 12
\end{eqnarray}

The distribution of these M giants in equatorial coordinates is shown in Fig. \ref{fig:radec_segue}. The excess of M giants near the Triangulum-Andromeda region (RA, DEC $\sim 30, 30$ deg) is the TriAnd overdensity that \cite{rocha04} discovered using a similar selection of M giants from 2MASS. The orange star indicates the location of the Segue 2 satellite, and the black box indicates the region we explore in the following sub-sections using SDSS/SEGUE spectra.

\subsection{A-type stars}

A-type stars consist mainly of Blue Horizontal Branch (BHB) and Blue Straggler (BS) stars located blue-ward of the MSTO. These stars are useful halo tracers owing to the limited contamination by other stellar populations (cf. dwarfs and giants at redder colours) and their relatively bright absolute magnitudes. We selected A-type stars from the SEGUE spectroscopic sample by applying the following colour and magnitude cuts:

\begin{eqnarray}
-0.5 < g-r < 0.1 \nonumber\\
0.5 < u-g < 2.0 \nonumber\\
18 < g < 20
\end{eqnarray}

BS stars dominate over BHB stars in this faint magnitude slice (\citealt{deason14}), and although they are $\sim 2$ mag fainter than BHB stars,  BS stars have proved a useful tool to study the prominent Sgr stream in the MW halo (\citealt{belokurov14}). The magnitude cut is applied to select BS stars with approximate distances of $10 \lesssim D/\mathrm{kpc} \lesssim 30$ (\citealt{deason11}), roughly coincident with the distance of the TriAnd overdensity (\citealt{majewski04}; \citealt{rocha04}; \citealt{sheffield14}).

In the right panel of Fig. \ref{fig:radec_segue} we show the spectroscopic SEGUE fields in the vicinity of the Segue 2 satellite and M31. The coloured points show the A-type stars selected with the above colour cuts. We only consider stars with $|b| > 20$ deg to avoid significant extinction corrections and disk contamination. The location of the Segue 2 satellite is shown with the orange star symbol. We also indicate the location of M31 and the PAndAS stream. The dashed line illustrates a great circle through the PAndAS stream, which coincides with some of the SEGUE fields.

The Galactocentric velocity distribution of A-type stars in the SEGUE fields are shown in Fig. \ref{fig:vlos_segue}. The location of the fields are colour-coded to match Fig. \ref{fig:radec_segue}. Significant velocity signatures coincident with the TriAnd overdensity are seen in the fields closest to the Segue 2 satellite, and roughly coincident with the great circle through the PAndAS stream. The fields at lower RA have less significant signatures, however there is a hint of the TriAnd velocity feature in the top right-hand panel. Note that in \S\ref{sec:msto} we show that a more prominent signature of the TriAnd overdensity is visible for MSTO stars in this region of the sky (see middle-left panels of Fig. \ref{fig:msto}). We perform a two-component (halo plus substructure) Gaussian fit to the velocity distributions in the fields near Segue 2 using the same maximum likelihood procedure as outlined in \S\ref{sec:splash_vel}. The double-Gaussian fit is shown in Fig. \ref{fig:vlos_segue} and the maximum likelihood parameters are given in Table \ref{tab:like}.

As noted previously by \cite{belokurov09}, the Segue 2 satellite has a very similar systemic velocity to TriAnd and is located at a comparable distance ($D \sim 30$ kpc). This, in addition to the PAndAS stream, is the second stream/satellite that has been found in close proximity (both spatially, and in velocity space) to the large TriAnd overdensity. It is plausible that these structures are all associated, and we discuss this further in \S\ref{sec:diss}.

\subsection{Giant Stars}

\begin{figure}
\begin{center}
\includegraphics[width=8.5cm, height=11.9cm]{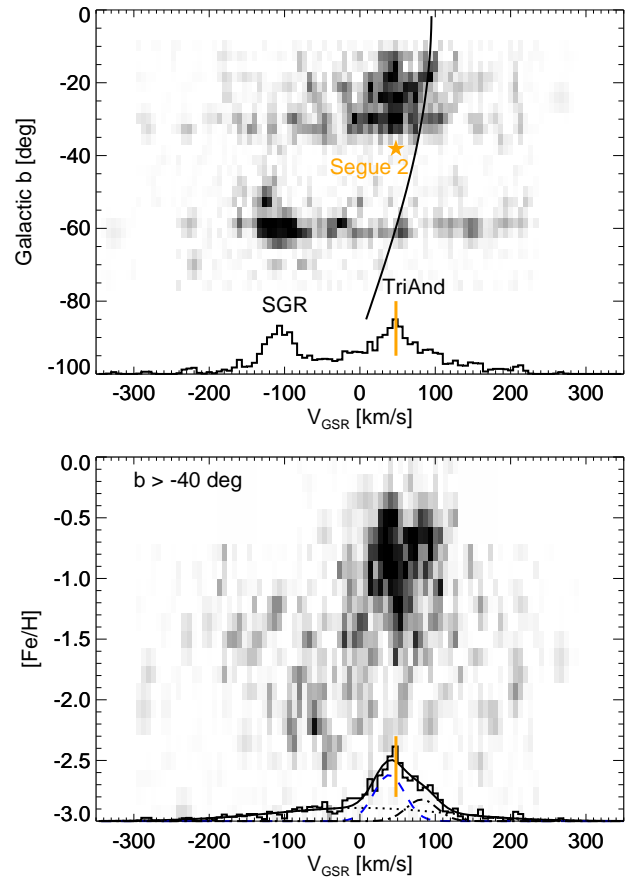}
\caption[]{The Galactocentric velocity distribution of giant stars selected from SEGUE with Galactic longitude $\ell \sim 150$ deg. \textit{Top panel:} The velocity distribution at different Galactic latitudes. The Sgr dwarf is prominent at latitudes away from the Galactic plane ($b < -40$ deg) and the TriAnd overdensity is significant at latitudes closer to the Galactic plane ($b > -40$ deg). The orange star symbol indicates the Galactocentric velocity and Galactic latitude of the Segue 2 satellite. The solid black line shows the velocity track of the thick disk as a function of Galactic latitude. 
\textit{Bottom panel:} The velocity distribution of SEGUE giants as a function of [Fe/H] metallicity. Only stars with $b > -40$ deg (and $b < -13$ deg) are shown to limit contamination from Sgr and thick disk stars with a similar velocity to TriAnd. The TriAnd feature is more metal-rich than the field halo giants, but has a wide distribution of metallicities. The solid line shows a triple Gaussian fit to the velocity distribution, which includes contributions from the halo, thick disk and TriAnd. The Gaussian parameters for the halo and thick disk are kept fixed in the fitting procedure. The blue dashed line shows the contribution from TriAnd.}
\label{fig:giants}
\end{center}
\end{figure}

Giants are useful tracers of halo populations owing to their bright absolute magnitudes, which allows relatively shallow spectroscopic surveys to probe out to large distances in the halo. Giant stars in SEGUE are selected using a similar selection to \cite{belokurov14}, but we include a wider colour range to include lower [Fe/H] stars:

\begin{eqnarray}
17 < g < 20  \nonumber \\
0.6 < g-i < 2.0 \nonumber\\
3.66 < \mathrm{log} (T_{\rm eff}/\mathrm{K}) < 3.76 \nonumber\\
1.5 < \mathrm{log}(g_s) < 3.8
\end{eqnarray}

\begin{table*}
\centering
\renewcommand{\tabcolsep}{0.2cm}
\renewcommand{\arraystretch}{1.2}
\begin{tabular} {l l l | c c c c c c}
 \hline
 \textbf{Survey} & \textbf{Substructure} & \textbf{Type} & $\mathbf{\ell, b}$ & $\mathbf{N_{\rm stars}}$ & $\mathbf{f_{\rm halo}(f_{\rm disk})}$  & $\mathbf{\langle V_{\rm GSR} \rangle}$ \textbf{[km s}$\mathbf{^{-1}}$\textbf{]} & $\mathbf{\sigma}$ \textbf{[km s}$\mathbf{^{-1}}$\textbf{]} & $\mathbf{\mathrm{ln}\left(\mathcal{L}_{\rm SUB}/\mathcal{L}_{\rm NO SUB}\right)}$\\
\hline
SPLASH & PAndAS stream & MSTO & 115.7, -15.9 & 44 & 0.69 & $67^{+8}_{-8}$ & $4^{+20}_{-4}$ & 16.1\\
SPLASH & TriAnd & MSTO & 123.1, -25.7  & 114 & 0.76 & $47^{+6}_{-8}$ & $12^{+13}_{-6}$ & 14.7\\
\hline
SEGUE & TriAnd & BS & 149.4, -27.8 & 78 & 0.42 & $36^{+6}_{-6}$ & $21^{+6}_{-5}$ & 29.4\\
SEGUE & TriAnd & BS & 172.3, -33.0 & 66 & 0.35 & $17^{+6}_{-8}$ & $19^{+12}_{-6}$ & 32.7\\
SEGUE & TriAnd & Giants & 150.0, -25.4 & 287 & 0.52(0.15) & $39^{+14}_{-6}$ & $20^{+14}_{-6}$ & 39.8\\
SEGUE & TriAnd & MSTO & 123.0, -32.5 & 506 & 0.80 & $51^{+7}_{-8}$ & $21^{+10}_{-7}$ & 24.7\\
\hline
\end{tabular}
\caption[]{The maximum likelihood parameters for the two/three-component (halo/disk plus substructure) Gaussian fits to the velocity distributions. The last column gives the log likelihood ratio for Gaussian distributions with ($\mathcal{L}_{\rm SUB}$) and without ($\mathcal{L}_{\rm NO SUB}$) a cold component.}
\label{tab:like}
\end{table*}

\begin{figure*}
  \centering
  \begin{minipage}{0.49\linewidth}
    \centering
    \includegraphics[width=8.5cm, height=6.8cm]{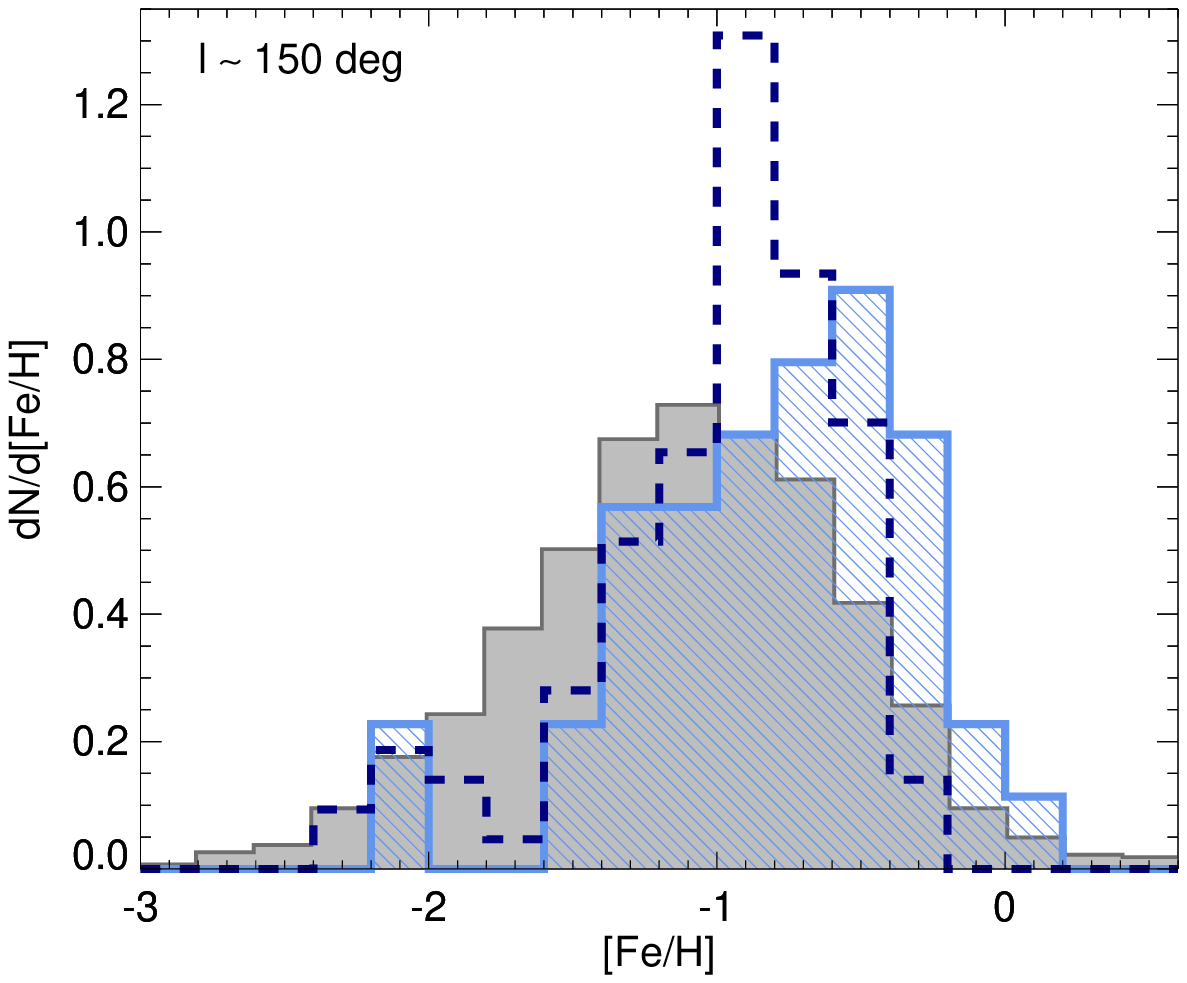}
  \end{minipage}\hfill
  \begin{minipage}{0.49\linewidth}
    \centering
    \includegraphics[width=8.5cm, height=6.8cm]{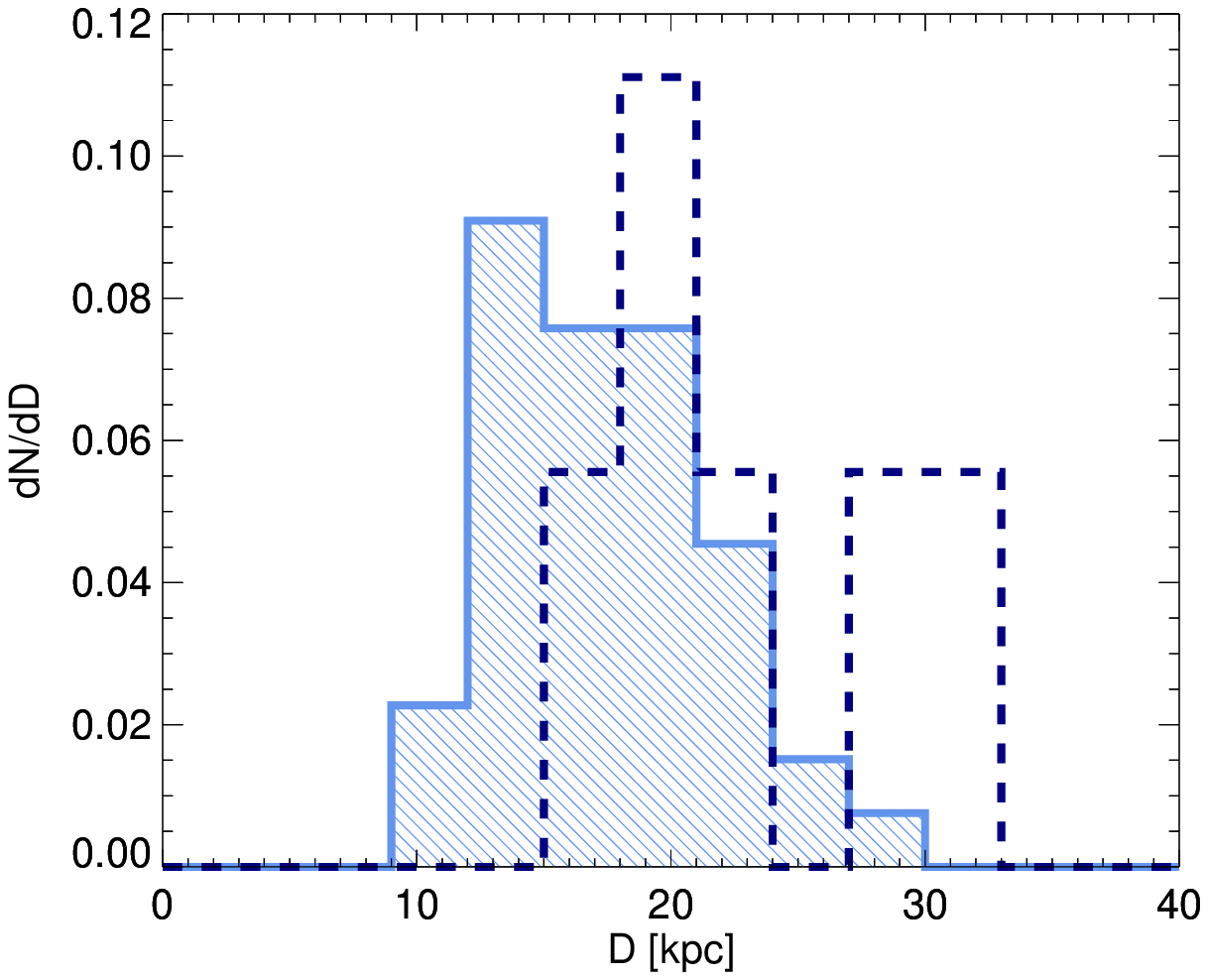}
  \end{minipage}
\caption[]{\textit{Left panel:} The metallicity distribution of ($N=2606$) A-type stars in SEGUE with $18 < g < 20$ and $20 < |b|/\mathrm{deg} < 40$ (grey filled histogram). The distribution for ($N=44$) SEGUE A-type stars in the TriAnd overdensity are shown with the light blue line-filled histogram. TriAnd is clearly more metal-rich than the overall halo population. The navy blue dashed line shows the metallicity distribution for ($N=107$) SEGUE giant stars in the TriAnd overdensity. The stars (both A-type and giants) in the TriAnd overdensity are selected from the $\ell \sim 150$ deg fields (see Table \ref{tab:like}) with $p_{\rm triand} > 0.5$. \textit{Right panel:} The approximate distances for the BS stars and giants in the TriAnd overdensity near $\ell \sim 150$ deg. The absolute magnitude calibration from \cite{deason11} is used for the ($N=44$) BS stars, assuming a metallicity of [Fe/H] $\sim -1.0$. We use the \cite{xue14} distance estimates for the giant stars; only ($N=6$) SEGUE giants which overlap with this catalog are shown in the figure.}
\label{fig:met_dist}
\end{figure*}

We focus on the SEGUE stripe with approximately constant $\ell = 150$ deg (see Fig. \ref{fig:radec_segue}), which contains the most obvious TriAnd signal and coincides with the Segue 2 satellite. In Fig. \ref{fig:giants} we show the velocity distribution of these giants as a function of Galactic latitude (top panel) and [Fe/H] metallicity (bottom panel). The black curve across the top panel is the predicted thick disk signature for dwarf contaminants in the sample. At higher latitudes, the thick disk has similar velocity to the TriAnd structure so it is difficult to distinguish these components. The Sgr stream is prominent at higher Galactic latitudes and negative Galactocentric velocities. In the bottom panel the velocity distribution is shown as a function of metallicity. In this panel, only stars with latitudes $b > -40$ deg (and $b < -13$ deg) are shown to limit confusion with the thick disk and Sgr. The velocity of the Segue 2 satellite is indicated and, as mentioned above,  is very similar to the TriAnd overdensity. The metallicity of TriAnd is more metal-rich than the field halo stars, but it has a very broad distribution with a tail to lower metallicities. In contrast, the thick disk distribution is clustered around relatively high metallicities ([Fe/H] $\sim -0.5$) with no significant low metallicity tail.

\subsection{Metallicity and distances of A-type stars and giants}

We study the metallicity distributions of the halo populations further in Fig. \ref{fig:met_dist}. In the left-hand panel the grey histogram shows the metallicity distribution of A-type stars selected in SEGUE with $20 < |b|/\mathrm{deg} < 40$. High probability TriAnd stars in the $\ell \sim 150$ deg SEGUE fields ($P_{\rm triand} > 0.5$) are shown with the line-filled light blue histogram. The TriAnd stars seem to be biased to higher metallicities. It is worth noting that the metallicity estimates for A-type stars likely suffer from significant systematic biases, and the [Fe/H] estimates from the SEGUE pipeline were not designed for these hot, A-type stars. Nonetheless, these metallicities should be sufficient to look at \textit{relative} differences between populations, and indeed, we find that the stars belonging to TriAnd tend to be more metal-rich than the field halo. 

The navy blue-dashed line shows the metallicity distribution of giant stars (with $\ell \sim 150$ deg) in SEGUE with high probability of belonging to TriAnd. Encouragingly, these metallicities agree very well with the A-type stars in TriAnd. We note that \cite{rocha04} and \cite{sheffield14} found that their samples of M-giants in TriAnd typically have metallicities of [Fe/H] $\sim -1.0$, in good agreement with our findings.
\begin{figure*}
  \centering
  \begin{minipage}{\linewidth}
    \centering
    \includegraphics[width=17cm, height=4.25cm]{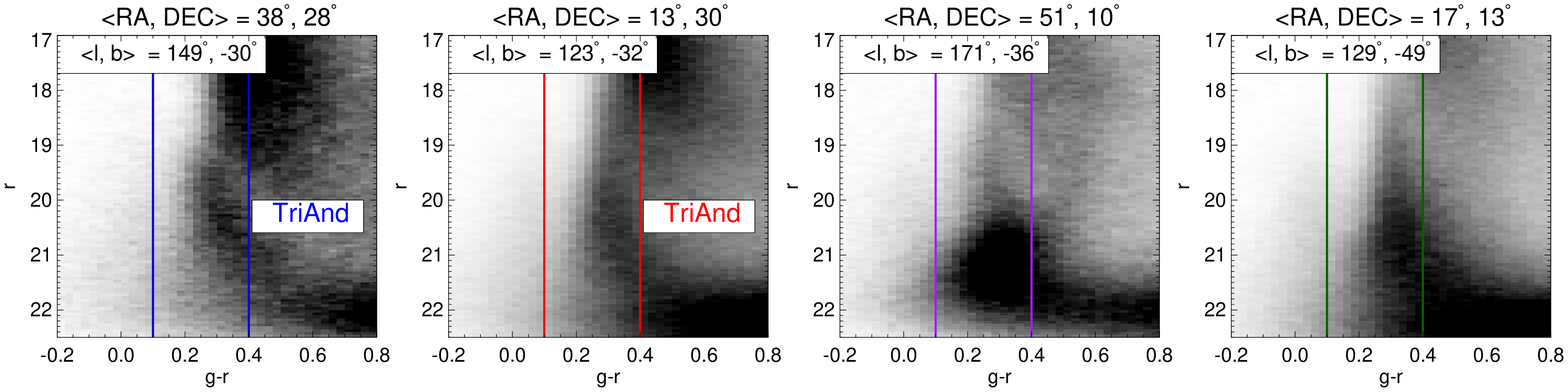}
  \end{minipage}\hfill
  \begin{minipage}{\linewidth}
    \centering
    \includegraphics[width=17cm, height=4.25cm]{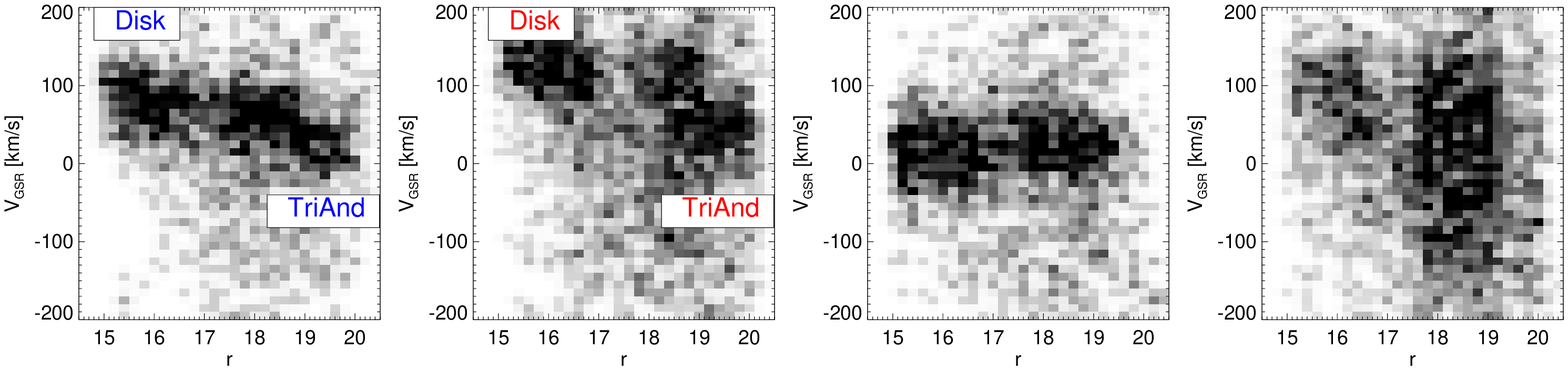}
  \end{minipage}
\caption[]{\textit{Top panels:} Colour-magnitude diagrams for stars in SDSS. Each panel is for different regions on the sky (average equatorial and Galactic coordinates are indicated) with the same colour coding as Fig. \ref{fig:radec_segue}. The solid lines show the $g-r$ colour region used to select MSTO stars. The TriAnd MS is clear in the left panels (cf. Fig. \ref{fig:washington_cmd}). The MSTO of the Sgr is evident in the third panel at $r \sim 21$. \textit{Bottom panels:} Galactocentric velocity against (r band) magnitude for MSTO stars. Each panel is for different regions on the sky as indicated in the top panels. The TriAnd overdensity at $V_{\rm GSR} \sim 40$ km s$^{-1}$ is clearly visible at $r \sim 19-20$. This velocity signature is distinct from the brighter disk stars.}
\label{fig:msto}
\end{figure*}

In the right-hand panel of Fig. \ref{fig:met_dist} we show the approximate distance distributions of the SEGUE stars likely related to TriAnd. For the A-type stars, we use the \cite{deason11} absolute magnitude calibration for BS stars assuming a metallicity of [Fe/H] $\sim -1.0$. The distances of the BS giant stars are shown by the light blue, line-filled region. We note that a small fraction of these stars may be BHBs, and these will have distances $D \gtrsim 30$ kpc. To estimate the distances of the giant stars, we cross match these stars with the SEGUE K-giant sample compiled by \cite{xue14}, who provide distance estimates to the giants. Only 6 (out of 111) stars are in the \cite{xue14} sample and their approximate distances are shown with the navy blue dashed line in the figure. Note that the small fraction of overlap between these samples is likely because the \cite{xue14} K-giants are biased towards lower-metallicity giants (e.g. less than 10\% of the sample have [Fe/H] $> -1.0$)

For both populations, BSs and giants, the distance estimates are highly uncertain, and we caution that these are only approximate values. However, it is encouraging that the BS stars and giants are in good agreement, with distances of $D \sim 20$ kpc, very similar to previous estimates for the TriAnd overdensity (\citealt{rocha04}; \citealt{majewski04}; \citealt{sheffield14}). We also note that our selection of A-type stars and giants identifies signatures of the Sgr stream, located at $D \sim 30$ kpc in this region of the sky, therefore giving further support to our distance calibration.

\subsection{Main Sequence Turn-Off Stars}
\label{sec:msto}

\begin{figure*}
\begin{center}
\includegraphics[width=18cm, height=9cm]{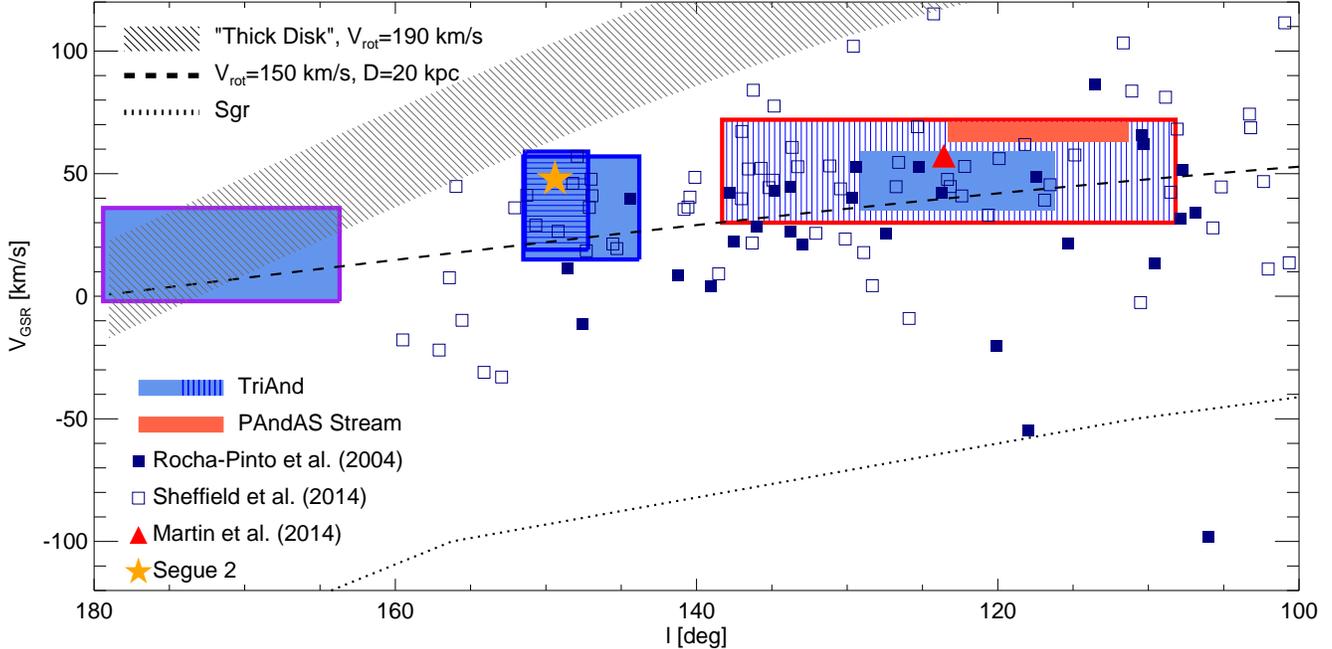}
\caption[]{Galactocentric velocity vs. Galactic longitude for TriAnd and its ``siblings''. The blue shaded (solid and line-filled) regions show the velocities of TriAnd in the SPLASH and SEGUE fields, and the red shaded region indicates the velocity of the PAndAS stream. The locations of the TriAnd fields for A-type stars in SEGUE are colour-coded (blue, purple solid lines) as in Fig. \ref{fig:radec_segue}. These shaded regions indicate the 1-$\sigma$ velocity distributions about the mean, and span the extent in Galactic longitude of each field. The horizontal line-filled region near the Segue 2 satellite shows the velocity distribution of TriAnd traced by giant stars.  The vertical line-filled region at lower Galactic longitude shows the velocity distribution of TriAnd traced by MSTO stars. The red outline follows the same colour-coding as Fig. \ref{fig:radec_segue} to indicate the location of the field. The dark blue squares show the velocities of the TriAnd M-giants compiled by \cite{rocha04} (filled) and \cite{sheffield14} (un-filled). The average velocity found by \cite{martin14} for the PAndAS stream is shown with the red triangle symbol, and the Segue 2 satellite (\citealt{belokurov09}) is shown with the orange star symbol. The dashed line indicates a circular orbit with $V_{\rm rot} = 150$ km s$^{-1}$, and the line-filled grey region shows a local thick disk-like population. The approximate projection of the Sgr stream orbit onto $V_{\rm GSR}, \ell$ space is indicated with the dotted line (\citealt{belokurov14}).}
\label{fig:master}
\end{center}
\end{figure*}

We end this section by looking at MSTO stars in the SDSS/SEGUE fields. This brings us right back to where we started in \S\ref{sec:splash_trash}, where we used MSTO stars in the SPLASH survey to probe the TriAnd overdensity. MSTO stars are fainter than A-type stars and giants, but there are many more of them. In fact, the high number density of MSTO stars has been exploited in previous work to map out streams and structures in the MW halo over large regions of the sky (e.g. \citealt{belokurov06a}).

In the top panels of Fig. \ref{fig:msto} we show CMDs for stars in SDSS imaging. Each panel shows the different regions of the sky that we probed in the previous subsections (see Fig. \ref{fig:radec_segue}). In the lower latitude fields ($ b \sim -30$ deg) the TriAnd MSTO is clear. A well defined MS for TriAnd is also apparent, which appears to be limited in distance as opposed to the blurry mess of halo stars spanning a large range of distances. In the left-hand panel ($(l, b) \sim (149, -30)$ deg) a gap between the more distant TriAnd MSTO and the disk population is evident. Furthermore, the TriAnd turn-off is slightly bluer, suggesting a metal-poorer population than the disk. 

The bottom panels of Fig. \ref{fig:msto} show the corresponding velocity distributions for the MSTO stars in each field (selected with $0.1< g-r < 0.4$). This allows us to distinguish between the brighter disk stars and the fainter stars in the MSTO of TriAnd. The velocity signatures of the thick disk and TriAnd are apparent in the two left-hand panels.  At brighter magnitudes, $16 \lesssim r \lesssim 19 $, there is a clear thick disk MSTO signature that has a gentle slope with $V_{\rm GSR}$ due to the projection of progressively more distant stars onto the line-of-sight. For $r \gtrsim 19$ the TriAnd signature appears clearly disconnected from the thick disk sequence with $V_{\rm GSR} \sim 40$ km s$^{-1}$. At Galactic longitudes close to $\ell \sim 180$ deg (third panel), the disk and halo signatures are difficult to disentangle as the disk crosses $V_{\rm GSR} =0$  km s$^{-1}$. Finally, the fourth panel displays a much broader halo population at fainter magnitudes ($\sigma \sim 100$ km s$^{-1}$), in good agreement with Fig. \ref{fig:vlos_segue}.

For the MSTO stars located in the second panel (indicated in red, $(\ell, b) \sim (123, -32)$ deg), we perform a two-component (halo plus substructure) fit to the velocity distribution of MSTO stars with $19 < r < 20$. The magnitude cut is included to remove the majority of disk stars. The resulting Gaussian parameters for the TriAnd feature are given in Table \ref{tab:like}. The results are in good agreement with our findings from the SPLASH survey (see \S\ref{sec:splash_vel}), which is located in a similar region on the sky (see red points in Fig. \ref{fig:radec_segue}).
\section{A fossil record of group infall?}
\label{sec:diss}

The results outlined in the previous sections suggest that at least three distinct substructures in the MW are associated, namely the large TriAnd overdensity, the PAndAS stream and the Segue 2 satellite. 

In Fig. \ref{fig:master} we give a summary of our results, showing Galactocentric velocity as a function of Galactic longitude for the various halo populations studied in this work. The shaded regions indicate the 1-$\sigma$ velocity distributions about the mean (see Table \ref{tab:like}) and span the extent in Galactic longitude of each field. Also shown are the Segue 2 satellite, the velocity measurement for the Pandas stream found in \cite{martin14}, and the TriAnd M-giant samples compiled by \cite{rocha04} and \cite{sheffield14}. Note that the velocity dispersion we find for the TriAnd features ($\sigma \sim 20$ km s$^{-1}$) is similar to the spread in velocities of the M-giant samples. For illustration, the dashed line shows the Galactocentric velocity profile for planar rotation at $D =20$ kpc (corresponding to Galactocentric distance, $r \sim 25$ kpc) with $V_{\rm rot}=150 $ km s$^{-1}$, and the line-filled region shows the approximate profile for a local thick disk population. The approximate projection of the Sgr stream orbit onto $V_{\rm GSR}, \ell$ space is also indicated with the dotted line (\citealt{belokurov14}).

It is clear that TriAnd, the PAndAS stream and Segue 2 may all follow very similar orbits, possibly even forming a ``ring-like'' structure with a roughly planar orbit ($V_{\rm rot} \approx V_{\rm \phi} \approx 150$ km s$^{-1}$), although without proper motion information we are unable to rule out different orbital configurations\footnote{The orbit is likely slightly inclined as the velocity signature doesn't seem to go through $V_{\rm GSR} =0$ km s$^{-1}$ at $l=180$ deg}. This suggests that these substructures fell into the MW halo as a group, with the large, metal-rich TriAnd structure dominating the group mass (i.e. the central) and the PAndAS stream and Segue 2 being satellites of TriAnd.

Wetzel, Deason \& Garrison Kimmel (2014, in prep.) recently showed that a significant fraction ($\sim 30$ percent) of ultra-faint dwarfs likely fell into the MW as satellites of larger dwarfs. This work also shows that group infall is more common for lower mass subhalos. Thus, it's perhaps unsurprising that the ``least massive galaxy'', Segue 2, can be likened to a sub-subhalo in the hierarchical structure formation paradigm. 

If Segue 2 was indeed born as a sub-system of a larger dwarf, then it may have undergone pre-processing by its original host's tidal field. This may explain why Segue 2 lies significantly off the metallicity-stellar mass relation that is closely followed by most MW dwarfs (\citealt{kirby13b}). However, it's also worth noting that, even if Segue 2 has no dynamical association with TriAnd, there is likely significant contamination from this large, metal-rich structure along the line-of-sight towards Segue 2. This could affect both the velocity dispersion measurements of the satellite, and the metallicity distribution of the apparent Segue 2 members, if not properly accounted for.

\section{Conclusions}
We have investigated the large, diffuse TriAnd overdensity using foreground MSTO star contamination in the SPLASH survey of M31, and using A-type stars, giants and MSTO stars from the SEGUE survey in the vicinity of the Segue 2 satellite. Our main conclusions are summarized as follows:

\begin{itemize}

\item Several of the SPLASH spectroscopic fields coincide with the recently discovered PAndAS stream, which is located at a similar distance ($D \sim 20$ kpc) and region of the sky to TriAnd. We find that the velocity of the stream is very similar to TriAnd ($V_{\rm GSR} \sim 50$ km s$^{-1}$), and there is tentative evidence for a velocity gradient in Galactic longitude along the PAndAS stream that is similar to the apparent gradient of the TriAnd overdensity over a much larger area of the sky.

\item At larger Galactic longitudes ($\ell \sim 150$ deg), the SEGUE fields allow us to probe the TriAnd overdensity in the vicinity of the Segue 2 satellite, which has previously been claimed to be coincident with this large substructure (\citealt{belokurov09}). Using A-type, giant and MSTO stars,  we find that the velocity of TriAnd stars in the vicinity of Segue 2 are very similar to the systemic velocity of the satellite.

\item The metallicity of TriAnd stars in the SEGUE survey are more metal-rich than the field halo. The Ca \textsc{ii} triplet lines from the SPLASH data have similar EWs in the PAndAS stream, TriAnd and the field halo, in apparent discord with our findings from SEGUE. However, these weak lines may be unable to distinguish between modest metallicity differences, especially for the relatively blue main-sequence stars under consideration.

\item We combine the results from the SPLASH and SEGUE surveys to map the TriAnd overdensity over $\sim 80$ deg in Galactic longitude. The velocity profile of the overdensity resembles a ``ring-like'' structure at $D \sim 20-30$ kpc. The coincidence of the PAndAS stream and Segue 2 satellite in positional and velocity space to TriAnd strongly suggests that they are all associated, and we advocate that these substructures may be a fossil record of group infall onto the MW halo. In this scenario,  the large, metal-rich TriAnd structure is the group central, whilst the PAndAS stream and Segue 2 are satellites of a Milky Way satellite.

\end{itemize}

\section*{Acknowledgments}
AJD thanks Allyson Sheffield and Andrew Wetzel for useful discussions. We thank the referee, Heidi Newberg, for providing useful comments on the paper. AJD is currently supported by NASA through Hubble Fellowship grant HST-HF-51302.01, awarded by the Space Telescope Science Institute, which is operated by the Association of Universities for Research in Astronomy, Inc., for NASA, under contract NAS5-26555. We thank the Aspen Center for Physics and the NSF Grant \#1066293 for hospitality during the conception of this paper. The research leading to these results has received funding from the European Research Council under the European Union's Seventh Framework Programme (FP/2007-2013) / ERC Grant Agreement n. 308024. Support for this work was provided by NASA through Hubble Fellowship grant 51273.01 awarded to K.M.G. by the Space Telescope Science Institute, which is operated by the Association of Universities for Research in Astronomy, Inc., for NASA, under contract NAS5-26555. Support for this work was provided by the National Science Foundation grants AST-1010039 (PI: Guhathakurta) and AST-1312863 (PI: Majewski).

\label{lastpage}

\end{document}